\documentclass{xpkas}

\def\year{2015} 
\def\month{January} 
\def\volume{00} 
\def\issue{0} 
\def\beginpage{1} 
\def\endpage{2} 
\def\received{October 31, 2014} 
\def\accepted{November 30, 2014} 
\date{Received \received ; accepted \accepted}

\newcommand\ion[2]{{#1}\,{\sc #2}} 

\def\apj{ApJ}

\def\apss{Ap\&SS}

\def\mnras{MNRAS}

\title{
Physical and Chemical Properties of \\Planetary Nebulae with WR-type Nuclei
}

\author[1]{Ashkbiz~Danehkar\thanks{E-mail: ashkbiz.danehkar@students.mq.edu.au}}
\author[2]{Roger~Wesson}
\author[3]{Amanda~Karakas}
\author[1,4]{Quentin~A.~Parker}

\affil[1]{Department of Physics and Astronomy, Macquarie University, Sydney, NSW 2109, Australia}
\affil[2]{European Southern Observatory, Alonso de C\'{o}rdova 3107, Casilla 19001, Santiago, Chile}
\affil[3]{Research School of Astronomy \& Astrophysics, Australian National University, Canberra ACT 2611, Australia}
\affil[4]{Australian Astronomical Observatory, PO Box 915, North Ryde, NSW 1670, Australia}

\def\corrauthor{
A.~Danehkar
}

\def\runningauthor{
A.~Danehkar et al. 
}

\def\runningtitle{
Chemical Properties of WR Planetary Nebulae
}

\def\keywords{
ISM: abundances  -- planetary nebulae: general -- 
stars: Wolf-Rayet 
}

\def\abstracttext{
We have carried out optical spectroscopic measurements of emission lines for a sample of Galactic planetary nebulae with Wolf-Rayet (WR) stars and weak emission-line stars (\textit{wels}). The plasma diagnostics and elemental abundance analysis have been done using both collisionally excited lines (CELs)
 and optical recombination lines (ORLs). 
It is found that the abundance discrepancy factors (ADF$\equiv$ORL/CEL) are closely correlated with the dichotomy between temperatures derived from forbidden lines and those from \ion{He}{i} recombination lines, implying the existence of H-deficient materials embedded in the nebula.
The H$\beta$ surface brightness correlations suggest that they might be also related to the nebular evolution.
}

\begin{document}
\pkashead 


\section{Introduction\label{wc2:sec:introduction}}

Observations of planetary nebulae (PNe) are used to determine the composition of the interstellar medium \citep[e.g.][]{Kingsburgh1994}, and to probe the physics of AGB stars  \citep[e.g.][]{Karakas2009}. Collisionally excited lines (CELs) have been extensively used to derive the abundances of heavy elements \citep[see e.g.][]{Kingsburgh1994,Liu2004b}. Alternatively, optical recombination lines (ORLs) have a much weaker dependence on temperature, thus resulting in more reliable abundance analyses. However, the abundances derived using the ORL method are systematically higher than those derived from CELs in PNe \citep[e.g.][]{Tsamis2004,Wesson2004,Wesson2005}. Previously, \citet{Peimbert1967} found a dichotomy between [O\,{\sc iii}] CEL and H~{\sc i} Balmer jump (BJ) temperatures with $T_{\rm e}$([O\,{\sc iii}]$)>T_{\rm e}$(BJ). Moreover, \citet{Wesson2005} found that $T_{\rm e}$([O\,{\sc iii}]$)>T_{\rm e}$(BJ$)>T_{\rm e}$(He\,{\sc i}$)>T_{\rm e}($O\,{\sc ii}), which was predicted by the two-phase models \citep{Liu2004b}, containing some cold ($T_{\rm e}\sim10^3$\,K) H-deficient materials, embedded in the warm ($T_{\rm e}\sim10^4$\,K) gas of normal abundances.

For this study, we carried out the optical integral field observations of a sample of PNe \citep[see][]{Danehkar2013b} using the Wide Field Spectrograph \citep[WiFeS;][]{Dopita2010} on the ANU 2.3 telescope. Our observations were carried out with the B7000/R7000 grating combination ($R\sim 7000$). 
We acquired series of bias, dome flat-field frames, twilight sky flats, arc lamp exposures, wire frames, spectrophotometric standard stars for flat-fielding, wavelength calibration, spatial calibration and flux calibration \citep[described in detail by][]{Danehkar2013a,Danehkar2014}. Suitable sky windows were selected from the science data for sky subtraction.  

\section{Plasma Diagnostics\label{wc2:sec:tempdens}}

Nebular electron temperatures $T_{\rm e}$ and densities $N_{\rm e}$ were obtained from the intrinsic intensities of CELs by solving level populations for an $n$-level ($\geqslant5$) atomic model using the \textsc{equib} code.

\begin{figure*}[t]
\centering
\includegraphics[width=70mm]{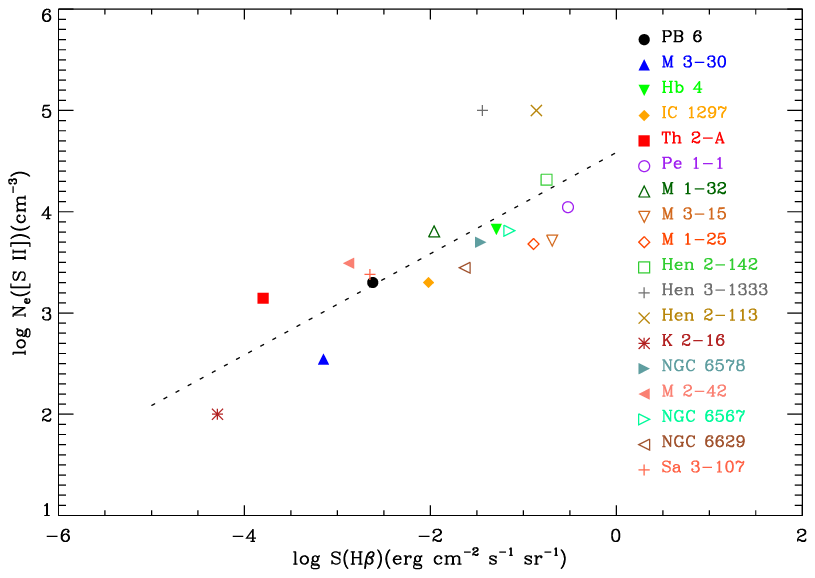}
\includegraphics[width=70mm]{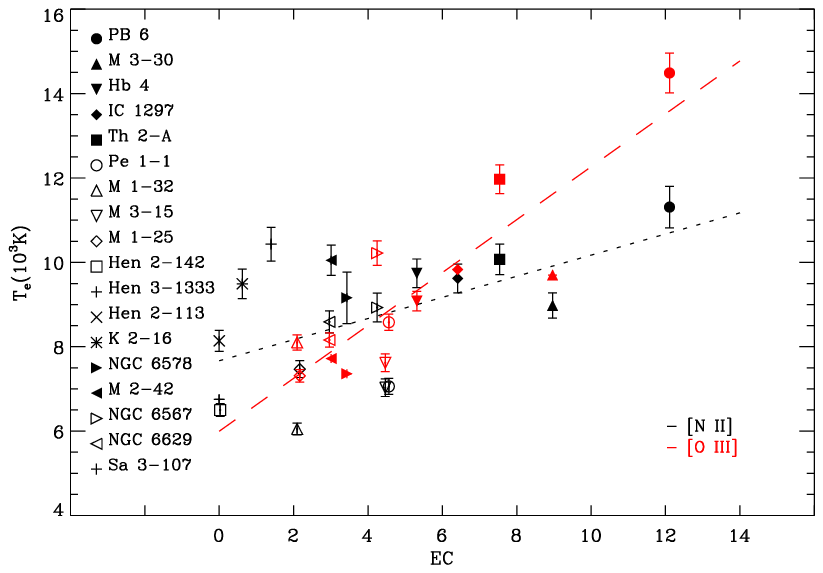}\\
\includegraphics[width=70mm]{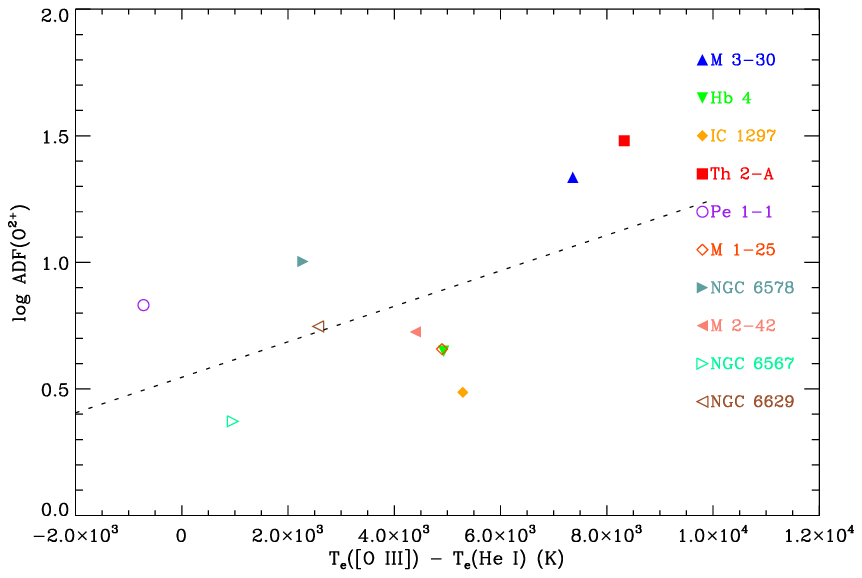}
\includegraphics[width=70mm]{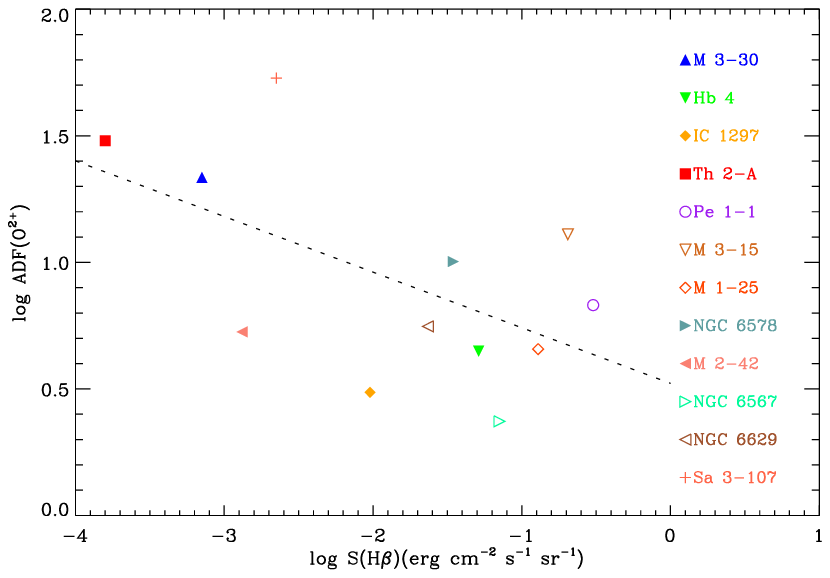}
\caption{Top-Left panel: The logarithmic electron density plotted against the logarithmic nebular surface brightness. 
Top-Right panel: Variation of the electron temperature along the excitation class (EC); the dotted and dashed lines for $T_{\rm e}$([N\,{\sc ii}]) and $T_{\rm e}$([O\,{\sc iii}]).
Bottom-Left panel: The difference between the electron temperatures and the helium temperatures plotted against the ADF
for O$^{2+}$. Bottom-Right panel: The ADF for O$^{2+}$ plotted against the logarithmic nebular surface brightness.
\label{wc2:fig1}
}
\end{figure*}

Fig.\,\ref{wc2:fig1} (top-left panel) shows the logarithmic electron density $\log N_{\rm e}$([S\,{\sc ii}]) plotted against the logarithmic intrinsic nebular H$\beta$ surface brightness. The dotted line represents a linear fit to the 18 PNe in our sample, which has a strong linear correlation:
$\log N_{\rm e}([{\rm S\,II}]) = (4.59 \pm 0.23)  +  (0.5 \pm  0.1)\,\log S({\rm H}\beta)$,
where $S({\rm H}\beta)$ is the dereddened nebular H$\beta$ surface brightness. It is seen that $S({\rm H}\beta) \propto N_{\rm e}^{2}$, in agreement with the theoretical relation by \citet{Odell1962}.

Fig.\,\ref{wc2:fig1} (top-right panel) plots $T_{\rm e}$ versus the excitation class \citep[EC;][]{Dopita1990}. A trend of increasing $T_{\rm e}$ with increasing EC is seen. A linear fit to $T_{\rm e}$([O\,{\sc iii}]) and as a function of EC yields: 
$T_{\rm e}([{\rm O\,III}]) = (5997 \pm 592)  + (626.84 \pm  100.57)\,{\rm EC}$.
The electron temperatures of high-excitation PNe are typically higher than low-excitation PNe, which can be explained by the radiation from the central stars. 

\section{Abundance Analysis\label{wc2:sec:abundances}}

We determined ionic abundances from CELs by solving the statistical equilibrium equations for each ion using the \textsc{equib} code, giving level population and line emmisivities for specified $T_{\rm e}$ and $N_{\rm e}$. We determined ionic abundances from ORLs for our sample where adequate observed lines were available.

Fig.\,\ref{wc2:fig1} (bottom-left panel) shows the logarithmic abundance discrepancy factor (ADF) for O$^{2+}$, defined as
$\log {\rm ADF}({\rm O}^{2+}) \equiv \log ({\rm O}^{2+}/{\rm H}^{+})_{\rm ORL}- \log({\rm O}^{2+}/{\rm H}^{+})_{\rm CEL}$
plotted against the difference between the [O\,{\sc iii}] forbidden-line and the He\,{\sc i} recombination-line temperatures
$\Delta T_{[{\rm O\,III}]} \equiv T_{\rm e}([{\rm O\, III}]) - T_{\rm e}({\rm He\, I})$.
A linear fit to the 10 PNe plotted in the figure yields: 
$\log {\rm ADF}({\rm O}^{2+})  =  (0.55 \pm 0.18) + (7.0 \pm 3.7) \times 10^{-5} 
 \times \Delta T_{[{\rm O\,III}]}({\rm K})$. 

In Fig.\,\ref{wc2:fig1} (bottom-right panel) we plot the O$^{2+}$/H$^{+}$ ADF as a function of the intrinsic nebular H$\beta$ surface brightness $\log S$(H$\beta$).  A linear fit to the 12 PNe plotted in 
the figure yields: 
$\log {\rm ADF}({\rm O}^{2+}) = (0.52 \pm 0.22) - (0.22 \pm 0.10)\,\log S({\rm H}\beta)$. 

\section{Conclusions\label{wc2:sec:discussions}}

In conclusion, there is a dependence of the nebular ORL/CEL ADFs upon the dichotomy between temperatures derived from forbidden lines and those from He~{\sc i} recombination lines, $T_{\rm e}($CELs$) - T_{\rm e}($He\,{\sc i}$)$, and the intrinsic nebular surface brightness, $\log S$(H$\beta$). 
It has been known that the ADFs are closely correlated with the difference between $T_{e}$([O\,{\sc iii}]) and $T_{e}$(BJ) \citep{Liu2004b,Tsamis2004,Wesson2005}. These correlations suggest that the observed ORLs may originate from cold ionized gas located in metal-rich clumps inside the diffuse warm nebula, but the origin of such inclusions is as yet unknown. The correlation between the nebular ADFs and the intrinsic H$\beta$ surface brightness found here is consistent with previous results \citep{Liu2004b,Tsamis2004}. This suggests that the abundance discrepancy might be related to the nebular evolution, and it is higher in old evolved PNe.


\acknowledgments

AD acknowledges a student travel assistance from the Astronomical Society of Australia (ASA) and a
travel grant from the International Astronomical Union (IAU).  




\begin{thebibliography}{}



\bibitem[{{Danehkar} {et~al}\mbox{.}(2013){Danehkar}, {Parker}, \&
  {Ercolano}}]{Danehkar2013a}
{Danehkar} A., {Parker} Q.~A., {Ercolano} B., 2013, \mnras, 434, 1513

\bibitem[{{Danehkar}(2014)}]{Danehkar2013b}
{Danehkar} A., 2014, PhD thesis, Macquarie University

\bibitem[{{Danehkar} {et~al}\mbox{.}(2014){Danehkar}, {Todt}, {Ercolano}, \&
  {Kniazev}}]{Danehkar2014}
{Danehkar} A., {Todt} H., {Ercolano} B., {Kniazev} A.~Y., 2014, \mnras, 439,
  3605
  

\bibitem[{{Dopita} \& {Meatheringham}(1990)}]{Dopita1990}
{Dopita} M.~A., {Meatheringham} S.~J., 1990, \apj, 357, 140


\bibitem[{{Dopita} {et~al}\mbox{.}(2010){Dopita}, {Rhee}, {Farage}, {McGregor},
  {Bloxham}, {Green}, {Roberts}, {Neilson}, {Wilson}, {Young}, {Firth},
  {Busarello}, \& {Merluzzi}}]{Dopita2010}
{Dopita} M. {et~al.}, 2010, \apss, 327, 245

\bibitem[{{Karakas} {et~al}\mbox{.}(2009){Karakas}, {van Raai}, {Lugaro},
  {Sterling}, \& {Dinerstein}}]{Karakas2009}
{Karakas} A.~I., {van Raai} M.~A., {Lugaro} M., {Sterling} N.~C., {Dinerstein}
  H.~L., 2009, \apj, 690, 1130
  
\bibitem[{{Kingsburgh} \& {Barlow}(1994)}]{Kingsburgh1994}
{Kingsburgh} R.~L., {Barlow} M.~J., 1994, \mnras, 271, 257

\bibitem[{{Liu} {et~al}\mbox{.}(2004){Liu}, {Liu}, {Barlow}, \&
  {Luo}}]{Liu2004b}
{Liu} Y., {Liu} X.-W., {Barlow} M.~J., {Luo} S.-G., 2004, \mnras,
  353, 1251

\bibitem[{{O'Dell}(1962)}]{Odell1962}
{O'Dell} C.~R., 1962, \apj, 135, 371

\bibitem[{{Peimbert}(1967)}]{Peimbert1967}
{Peimbert} M., 1967, \apj, 150, 825



\bibitem[{{Tsamis} {et~al}\mbox{.}(2004){Tsamis}, {Barlow}, {Liu}, {Storey}, \&
  {Danziger}}]{Tsamis2004}
{Tsamis} Y.~G., {Barlow} M.~J., {Liu} X.-W., {Storey} P.~J., {Danziger} I.~J.,
  2004, \mnras, 353, 953

\bibitem[{{Wesson} \& {Liu}(2004)}]{Wesson2004}
{Wesson} R., {Liu} X.-W., 2004, \mnras, 351, 1026

\bibitem[{{Wesson} {et~al}\mbox{.}(2005){Wesson}, {Liu}, \&
  {Barlow}}]{Wesson2005}
{Wesson} R., {Liu} X.-W., {Barlow} M.~J., 2005, \mnras, 362, 424


\end{thebibliography}
\end{document}